\documentclass[preprint,amsmath,amssymb,11pt,english,aps,showpacs,showkeys]{revtex4}
\usepackage{graphicx}
\usepackage{graphics}
\usepackage{amsmath}
\usepackage{dcolumn}
\usepackage{amssymb}
\usepackage{bm}
\usepackage[latin1]{inputenc}

\begin{document}
\title{Relativistic Landau  quantization for a neutral particle}
\author{K. Bakke and C. Furtado}
\email{kbakke@fisica.ufpb.br,furtado@fisica.ufpb.br} 
\thanks{phone number +558332167534}
\affiliation{Departamento de F\'{\i}sica, Universidade Federal da Para\'{\i}ba, Caixa Postal 5008, 58051-970, Jo\~{a}o Pessoa, PB, Brazil}

\begin{abstract}
In this contribution we study the Landau levels arising within the relativistic quantum dynamics of a neutral particle which possesses a permanent magnetic dipole moment interacting with an external electric field. We consider the Aharonov-Casher coupling of magnetic dipole to the electric field to investigate an an analog of Landau quantization in this system and solve the Dirac equation for two different field configurations.  The eigenfunctions and eigenvalues of Hamiltonian in both cases are obtained.
\end{abstract}

\maketitle

\section{Introduction}

The Landau quantization \cite{landau} is known in the literature as the quantization of cyclotron orbits for a charged particle motion when this particle interacts with an external magnetic field in the non-relativistic regime. The Landau quantization in non-relativistic regime is also discussed in the cases of other physical systems such as Bose-Einstein condensate \cite{paredes1,paredes2}, different two-dimensional surfaces \cite{comtet,grosche,dunne} and quantum Hall effect \cite{prange}. For the relativistic dynamics of a charged particle, the Landau quantization was first discussed by Jackiw \cite{rl1} and by Balatsky \textit{et al} \cite{rl2}. Other studies of the relativistic Landau quantization were carried out in the cases of a quantum Hall effect \cite{rl3}, a spin nematic state \cite{rl4} and a finite temperature  \cite{rl5}. 

In the case of the non-relativistic dynamics of a neutral particle, Ericsson and Sj\"oqvist\cite{er} have used the Aharonov-Casher(AC) ~\cite{ac} interaction to generate an analogue of the Landau quantization in a system of a neutral particle possessing a  magnetic dipole moment interacting with an external electric field. The AC effect is the effect reciprocal to the Aharonov-Bohm (AB) effect \cite{ab}. Other important topological effect is the dual AC effect, which is known as He-McKellar-Wilkens (HMW) effect \cite{hm,w}. Within the HMW effect, the phase shift is generated by the interaction between an external magnetic field and the electric dipole moment of the neutral particle. Thus, considering the HMW effect, Ribeiro \textit{et al} \cite{2} have studied the analog of Landau quantization for a neutral particle which possesses a permanent electric dipole moment, and Furtado \textit{et al} \cite{3} have investigated the Landau quantization for induced electric dipoles, being motivated by the paper of Wei \textit{et al} \cite{rl6}. Within the framework of the noncommutative quantum mechanics, Passos \textit{et al} \cite{fur2} have studied geometric phases for neutral particles, and the Landau quantization in this case was studied by Ribeiro \textit{et al} \cite{fur3}. Recently, The Landau quantization was investigated in a system of an atom in the presence of an external electric and a inhomogeneous magnetic field by Basu \textit{et al} \cite{basu}.  In the context of the Lorentz-symmetry violation, Ribeiro \textit{et al} \cite{fur4} investigated the arising of geometric phases for neutral particles and Passos \textit{et al} \cite{fur5} investigated the analog of Landau levels.  

In this paper, we construct an analog of the relativistic Landau quantization for a neutral particle with a permanent magnetic dipole moment coupled to an external electric field. We develop the relativistic analog of the Landau quantization in the symmetric gauge and in the Landau gauge. Finally, we take the non-relativistic limit of the relativistic energy levels and compare it with the respective non-relativistic cases.

This paper is organized as follows. In section II, we study the relativistic analog of the Landau quantization of neutral particle with a permanent magnetic dipole moment when the electric field is given in the symmetric gauge. In section III, we study the relativistic analog of the Landau quantization for the neutral particle when the Landau gauge for the electric field is imposed. In section IV, we present our conclusions.

\section{Relativistic analogue of the Landau Levels: the symmetric gauge}

In this section, we study the relativistic Landau levels arising within the relativistic dynamics of neutral particles with permanent magnetic dipole moment coupled with an external electric field. The quantum dynamics corresponding to this interaction arises due to the introduction of the non-minimal coupling into the Dirac equation \cite{bjd}
\begin{eqnarray}
i\gamma^{\mu}\,\partial_{\mu}\rightarrow i\gamma^{\mu}\,\partial_{\mu}+\frac{\mu}{2}\Sigma^{\mu\nu}\,F_{\mu\nu},
\label{5.0}
\end{eqnarray}
where $\mu$ is a magnetic dipole moment of the neutral particle and $\Sigma^{\mu\nu}=\frac{i}{2}\left[\gamma^{\mu},\gamma^{\nu}\right]$. The $\gamma^{\mu}$ matrices are the Dirac matrices in the flat spacetime, \textit{i.e.},
\begin{eqnarray}
\gamma^{0}=\hat{\beta}=\left(
\begin{array}{cc}
1 & 0 \\
0 & -1 \\
\end{array}\right);\,\,\,\,\,\,
\gamma^{i}=\hat{\beta}\,\hat{\alpha}^{i}=\left(
\begin{array}{cc}
 0 & \sigma^{i} \\
-\sigma^{i} & 0 \\
\end{array}\right);\,\,\,\,\,\,\Sigma^{i}=\left(
\begin{array}{cc}
\sigma^{i} & 0 \\
0 & \sigma^{i} \\	
\end{array}\right),
\label{5.0a}
\end{eqnarray}
with $\gamma^{\mu}\gamma^{\nu}+\gamma^{\nu}\gamma^{\mu}=-2\eta^{\mu\nu}$, $\vec{\Sigma}$ being the spin vector and $\sigma^{i}$ the Pauli matrices. The tensor $\eta^{\mu\nu}=diag(- + + +)$ is the Minkowsky tensor and the indices $(i,j,k=1,2,3)$. 
The Dirac equation for this system becomes 
\begin{eqnarray}
i\gamma^{0}\frac{\partial\Psi}{\partial t}+i\gamma^{1}\frac{\partial\Psi}{\partial x}+i\gamma^{2}\frac{\partial\Psi}{\partial y}+i\gamma^{3}\frac{\partial\Psi}{\partial z}+i\mu\,\vec{\alpha}\cdot\vec{E}\Psi-m\Psi=0.
\label{5.1}
\end{eqnarray}

Now let us configure the electric field in a way within which the necessary conditions pointed out in the reference \cite{er} for the field-dipole configuration are satisfied, to obtain a Landau quantization. 
The field configuration in the relativistic regime must satisfy the same conditions pointed out in \cite{er} for the non-relativistic dynamics, \textit{i.e},
\begin{eqnarray}
\frac{\partial\vec{E}}{\partial t}=0;\,\,\,\,\,\,\,\,\,\,\vec{\nabla}\times\vec{E}=0,
\label{2.7}
\end{eqnarray}
and also the condition of absence of torque on the magnetic dipole.
We choose an electric field satisfying the field condition given in (\ref{2.7}), in the following form
\begin{eqnarray}
\vec{E}=\frac{\lambda}{2}\,\left(x,y,0\right),
\label{2.6}
\end{eqnarray}
with $\lambda$ is a linear density charge and $x$ and $y$ can be measured with respect to the symmetry axis of the cylinder. This field configuration can be obtained inside of two cylindrical shells.

Using the field configuration given in the expression (\ref{2.6}), we study the relativistic Landau-Aharonov-Casher quantization which occurs within the relativistic dynamics of the neutral particle which has a permanent magnetic dipole moment interacting with an external electric field. 
Taking $\Psi=e^{-i\mathcal{E}t}\,\psi\left(x\right)$ and the external electric field (\ref{2.6}), we can rewrite the Dirac equation (\ref{5.1}) in the form
\begin{eqnarray}
\mathcal{E}\psi=m\hat{\beta}\psi+\hat{\alpha}^{1}\left(\hat{p}_{x}+i\frac{\mu\lambda}{2}x\,\hat{\beta}\right)\psi+\hat{\alpha}^{2}\left(\hat{p}_{y}+i\frac{\mu\lambda}{2}y\,\hat{\beta}\right)\psi+\hat{\alpha}^{3}\hat{p}_{z}\psi.
\label{5.3}
\end{eqnarray}
We can rewrite the Eq.(\ref{5.3}) in the following form
\begin{eqnarray}
\mathcal{E}\psi=m\hat{\beta}\psi+\hat{\alpha}^{1}\left(\hat{p}_{x}+i\mu A^{AC}_{x}\right)\psi+\hat{\alpha}^{2}\left(\hat{p}_{y}+i\mu A^{AC}_{y}\right)\psi+\hat{\alpha}^{3}\hat{p}_{z}\psi.
\label{5.3.1}
\end{eqnarray}
Notice that the components of the vector $A^{AC}_{x}=\frac{\lambda}{2}x\,\hat{\beta}$ and $A^{AC}_{y}=\frac{\lambda}{2}y\,\hat{\beta}$ play the role of a vector potential. This effective potential vector was denominated as Aharonov-Casher potential vector. Note that this field configuration generates a symmetric gauge configuration for the AC potential. In this case, we have an effective uniform magnetic field given by 
\begin{eqnarray}
\vec{B}_{AC}=\vec{\nabla}\times\vec{\Xi}=\mu\,\lambda\,\hat{z}.
\label{2.8}
\end{eqnarray}
Now, we meet an effective problem of solving the Dirac equation for a particle with an effective charge $\mu$ minimally coupled to the Aharonov-Casher gauge potential $\vec{A^{AC}}$. This problem is similar to the relativistic problem of solving the Dirac equation for the electric charge placed in an uniform magnetic field.

Using the cylindrical symmetry of the system, we suggest the solution of the Dirac equation (\ref{5.3}) to be in the form 
\begin{eqnarray}
\psi\left(x\right)=Ce^{il\varphi}\,e^{ikz}\,\left(
\begin{array}{c}
R_{1}\left(\rho\right)\\
R_{2}\left(\rho\right)\\	
\end{array}\right).
\label{5.4}
\end{eqnarray}
Where $C$ is a constant spinor amplitude.
Substituting the solution (\ref{5.4}) into the equation (\ref{5.3}), suggesting that the magnetic dipole moments are parallel to the $z$-axis, we arrive at the following equations for $R_{1}\left(\rho\right)$ and $R_{2}\left(\rho\right)$
\begin{eqnarray}
\left[\frac{d^{2}}{d\rho^{2}}+\frac{1}{\rho}\frac{d}{d\rho}-\frac{l^{2}}{\rho^{2}}-\frac{\mu^{2}\lambda^{2}}{4}\rho^{2}-l\mu\lambda-\mu\lambda+\left(\mathcal{E}^{2}-m^{2}-k^{2}\right)\right]\left(
\begin{array}{c}
R_{1}\left(\rho\right)\\
R_{2}\left(\rho\right)\\	
\end{array}\right)=0.
\label{5.5}
\end{eqnarray}
Now, we make a change of variables 
\begin{eqnarray}
\tau=\frac{\mu\lambda}{2}\,\rho^{2},
\label{5.6}
\end{eqnarray}
so the equation (\ref{5.5}) becomes
\begin{eqnarray}
\left[\tau\,\frac{d^{2}}{d\tau^{2}}+\frac{d}{d\tau}+\left(\beta-\frac{l^{2}}{4\tau}-\frac{\tau}{4}\right)\right]\left(
\begin{array}{c}
R_{1}\left(\rho\right)\\
R_{2}\left(\rho\right)\\	
\end{array}\right)=0,
\label{5.7}
\end{eqnarray}
with the new parameter $\beta$ is  given by
\begin{eqnarray}
\beta=\frac{1}{2\mu\lambda}\left(\mathcal{E}^{2}-m^{2}-k^{2}\right)-\frac{l}{2}-\frac{1}{2}.
\label{5.8}
\end{eqnarray}

The solution for the equation (\ref{5.7}) has the form
\begin{eqnarray}
R\left(\tau\right)_{1,2}=e^{-\frac{\tau}{2}}\,\tau^{\frac{\left|l\right|}{2}}\,F\left(-n,\left|l\right|+1;\tau\right),
\label{5.9}
\end{eqnarray}
and the relativistic energy levels which can be treated as the relativistic analogs of the Landau levels for a neutral particle with a permanent magnetic dipole moment are
\begin{eqnarray}
\mathcal{E}^{2}_{n}=m^{2}+k^{2}+2\mu\lambda\left(n+\frac{\left|l\right|}{2}+\frac{l}{2}+1\right),
\label{5.10}
\end{eqnarray}
where $n=0,\pm1,\pm2,\pm3,...$, which shows us that the relativistic analogs of the Landau levels are infinitly degenerated. The radial eigenfunctions are
\begin{eqnarray}
R_{n,l}\left(\rho\right)=\left(\frac{\mu\lambda}{2}\right)^{\frac{\left|l\right|}{2}}\,e^{-\frac{\mu\lambda}{4}\rho^{2}}\,\rho^{l}\,F\left(-n,\left|l\right|+1;\frac{\mu\lambda}{2}\rho^{2}\right).
\label{5.10a}
\end{eqnarray}
Using the condition of absence of torque on the dipole, we find that eigenvalues are given by
\begin{eqnarray}
\mathcal{E}^{2}_{n}=m^{2}+2\mu\lambda\left(n+\frac{\left|l\right|}{2}+\frac{l}{2}+1\right),
\label{5.10}
\end{eqnarray}
where $n=0,\pm1,\pm2,\pm3,...$

At this moment, we want to analyze the nonrelativistic limit of the relativistic analog of the Landau levels. Let us write the expression for the relativistic analogs of the Landau levels (\ref{5.10}) in the following form
\begin{eqnarray}
\mathcal{E}_{n}=m\,\sqrt{1+\frac{2\mu\lambda}{m^{2}}\left(n+\frac{\left|l\right|}{2}+\frac{l}{2}+1\right)}.
\label{5.13}
\end{eqnarray}
Applying the Taylor expansion up to the first order term in (\ref{5.13}), one finds the following non-relativistic analogs of the Landau levels
\begin{eqnarray}
\mathcal{E}_{n}\approx m+\frac{\mu\lambda}{m}\left[n+\frac{\left|l\right|}{2}+\frac{l}{2}+1\right],
\label{5.14}
\end{eqnarray}
where the first term in the right-hand side of the equation is the rest mass of the neutral particle. The last terms of (\ref{5.14}) correspond to the non-relativistic analog of the Landau levels obtained in \cite{er}, where these non-relativistic energy levels are infinitely degenerated. 

\section{relativistic analogue of the Landau levels in Landau gauge}
$ $

In this section, we study the relativistic analog of the Landau levels for a neutral particle which possesses a permanent magnetic dipole moment interacting with an external electric field given for a uniformly charged plane, where in the $x$-direction the plane is finite, whereas in the $y$ and $z$-directions it is infinite. This field configuration corresponds to the following electric field 
\begin{eqnarray}
\vec{E}=\lambda\left(x,0,0\right),
\label{6.1}
\end{eqnarray}
where this field configuration satisfies the conditions (\ref{2.7}) which is established in the reference \cite{er}. The Dirac equation for this field configuration becomes
\begin{eqnarray}
\mathcal{E}\psi=m\hat{\beta}\psi+\hat{\alpha}^{1}\left(\hat{p}_{x}+i\mu\lambda\,x\,\hat{\beta}\right)\psi+\hat{\alpha}^{2}\hat{p}_{y}\psi+\hat{\alpha}^{3}\hat{p}_{z}\psi,
\label{6.2}
\end{eqnarray}
where we have used the ansatz $\Psi=e^{-i\mathcal{E}t}\,\psi\left(x,y,z\right)$ and the external electric field (\ref{6.1}).
The Eq. (\ref{6.2}) can be rewritten in the following form
\begin{eqnarray}
\mathcal{E}\psi=m\hat{\beta}\psi+\hat{\alpha}^{1}\left(\hat{p}_{x}+i\mu
A^{AC}_{x}\right) )\psi+\hat{\alpha}^{2}\hat{p}_{y}\psi+\hat{\alpha}^{3}\hat{p}_{z}\psi,
\label{6.2.1}
\end{eqnarray}
where the gauge potential is chosen to look like $A^{AC}_{x}=\lambda\,x\,\hat{\beta}$. Note that the field configuration (\ref{6.1}) produces an Aharonov-Casher vector potential similar to the vector potential for the Landau levels in the Landau gauge.

 Let us take the ansatz for the solution of the equation (\ref{6.2}) in the following form
\begin{eqnarray}
\psi\left(x,y,z\right)=Ce^{ip_{y}y}\,e^{ikz}\,\left(
\begin{array}{c}
R_{1}\left(x\right)\\
R_{2}\left(x\right)\\	
\end{array}\right).
\label{6.3}
\end{eqnarray}
where $C$ is a constant spinor.
Substituting the solution (\ref{6.3}) into the equation (\ref{6.2}) and supposing that the magnetic dipole moments are parallel to the $z$-axis, we obtain the following equations for $R_{1}\left(x\right)$ and $R_{2}\left(x\right)$:
\begin{eqnarray}
\left(\mathcal{E}^{2}-m^{2}-k^{2}-\mu\lambda\right)\,R_{1,2}\left(x\right)=\left[\hat{p}_{x}^{2}+\mu\lambda\left(x+\frac{p_{y}}{\mu\lambda}\right)^{2}\right]\,R_{1,2}\left(x\right).
\label{6.4}
\end{eqnarray}
The right-hand-side of the equation (\ref{6.4}) is an analog of the one-dimensional harmonic oscillator Hamiltonian with the minimum  at $x_{0}=-\frac{p_{y}}{\mu\lambda}$. Thus, the relativistic energy levels for the neutral particle interacting with the external electric field (\ref{6.1}) are
\begin{eqnarray}
\mathcal{E}^{2}_{n}=m^{2}+k^{2}+2\mu\lambda\left(n+\frac{1}{2}\right) + \mu\lambda,
\label{6.5}
\end{eqnarray}
where $n=0,\pm1,\pm2,...$. If we impose the condition of absence of torque on the dipole, we obtain the following eigenvalues
\begin{equation}
 \mathcal{E}^{2}_{n}=m^{2}+2\mu\lambda\left(n+\frac{1}{2}\right) + \mu\lambda,
\label{6.5.1}
\end{equation}
Hence, we find that the energy levels given in (\ref{6.5.1}) are the relativistic analogs of the Landau levels for a neutral particle with a permanent magnetic dipole moment interacting with a field configuration given in the Landau gauge. The eigenfunctions are
\begin{eqnarray}
\psi\left(x,y,z\right)=Ce^{ip_{y}y}\,e^{ikz}\,R_{1,2}\left(x+\frac{p_{y}}{\mu\lambda}\right),
\label{6.5a}
\end{eqnarray}
with $R_{1,2}\left(x+\frac{p_{y}}{\mu\lambda}\right)=H_{n}(x+\frac{p_{y}}{\mu\lambda})$ being the Hermite polynomials.

The non-relativistic values for the analogs of Landau levels for the neutral particle in the Landau gauge can be obtained via the Taylor expansion. We rewrite the expression (\ref{6.5.1}) in the form
\begin{eqnarray}
\mathcal{E}_{n}=m\sqrt{1+2\frac{\mu\lambda}{m^{2}}\left(n+\frac{1}{2}\right)+\frac{\mu\lambda}{m^{2}}}.
\label{6.6}
\end{eqnarray}
Thus, expanding up to the first order, we have 
\begin{eqnarray}
\mathcal{E}_{n}=m+\frac{\mu\lambda}{m}\left(n+\frac{1}{2}\right) +\frac{\mu\lambda}{2m},
\label{6.7}
\end{eqnarray}
where the first term corresponds to the rest mass of the neutral particle, the last terms corresponds to the non-relativistic analogue of the Landau levels with the field configuration given in the Landau gauge \cite{er}.

\section{conclusions}
$ $ 
We studied the relativistic analog of the Landau quantization for a neutral particle with a permanent magnetic dipole moment interacting with an external electric field. We suggested that the field configuration which generated the Landau quantization in the relativistic dynamics of this neutral particle must satisfy the same conditions imposed by Ericsson and Sj\"oqvist \cite{er} in the non-relativistic dynamics.
For the relativistic dynamics of the neutral particle with permanent magnetic dipole moment interacting with an external electric field, the relativistic analog of the Landau levels arises naturally. At the end, we investigated the non-relativistic limit of the energy levels obtained in the expression (\ref{5.10}) through the Taylor expansion to the first order. We obtained, in the non-relativistic limit, the same expression for the non-relativistic analog of the Landau levels in \cite{er}.

We claim that we can obtain the results of quantum dynamics of neutral particle with a permanent electric dipole moment interacting with an external magnetic field via the He-Mckellar-Wilkens coupling \cite{hm,w}. We carried out this study employing the duality transformation in the equations of motions of the Landau-Aharonov-Casher problem in the presence of a topological defect and obtained the He-Mckellar-Wilkens quantization for a neutral particle in the presence of a topological defect. The equations of motion for the HMW case can be easily obtained through the changing $\mu\rightarrow d$ and $\vec{E}\rightarrow\vec{B}$, where $d$ is the electric dipole moment of the neutral particle and $\vec{B}$ is the magnetic field.

\acknowledgments{Authors are grateful to A. Yu. Petrov for some criticism on the manuscript. This work was partially supported by PRONEX/FAPESQ-PB, FINEP, CNPq and CAPES/PROCAD.}


\end{document}